%% file: Fewbody4.tex
\documentclass[seceq]{ptptex}

\usepackage{graphicx}
\include{defs}

\nofigureboxrule                     
\notypesetlogo

\title{
Baryon-Baryon Interaction in the Quark Cluster Model%
}
\author{
Makoto \textsc{Oka}%
}
\inst{
Department of Physics, Tokyo Institute of Technology\\
Meguro, Tokyo 152-8551 Japan\\
Email:{oka@th.phys.titech.ac.jp}
}

\abst{
The quark cluster model approach to the baryon-baryon interaction is 
reviewed and recent application to the  charge symmetry breaking in 
nuclear force is discussed.
}

\begin{document}

\maketitle

\section{Introduction}

According to quantum chromodynamics (QCD), hadrons are bound states of 
quarks and gluons, and their interactions are governed by the 
non-Abelian gauge interaction.  
Their structures are potentially very complex as QCD is strongly interacting 
in the hadronic energy region. 
Indeed, we have many reasons to believe that the QCD vacuum (ground 
state) has nontrivial structure, such as chiral symmetry breaking, 
nontrivial topology.
On the other hand, the low energy excitations of mesons and baryons show 
regular spectrum, which indicates their structures are rather simple.
In terms of the Fock space decomposition, dominant component of a 
meson or a baryon seems to be simply $q\bar q$ or $qqq$.

Chiral symmetry breaking is a key to resolve this seemingly 
contradictory situation.  That is, spontaneous chiral symmetry 
breaking induces a gap in the  
quark spectrum, so that the effective quark, called constituent quark, 
acquires a mass of order 300 MeV.  As this mass gap reduces the mixing 
of higher Fock components, or mixing of extra $q\bar q$, the leading Fock  
component becomes dominant.  
This is the basis of the quark model, in which hadrons can be treated as 
few-body quantum systems and various techniques developed for few-body 
quantum systems may be applied. 

One fruitful application of the quark model is the quark cluster 
model (QCM) approach to the baryon-baryon interactions.
The QCM, originally developed for the nucleon-nucleon interaction, 
successfully explained the origin of the strong short-range 
repulsion{\cite{QCM}}.
Later it was applied to other baryons, such as the $\Delta$ resonance, 
and the hyperons, $\Lambda$, $\Sigma$, and $\Xi$.  
Short range parts of such baryon-baryon interactions, which had 
not been well studied experimentally, have been predicted. 
For instance, the $H$ dibaryon\cite{Jaffe} is predicted to be a bound state of 
$\Lambda\Lambda$ in QCM\cite{H}, which motivated extensive dibaryon searches.

Realistic baryon-baryon interactions have been generated in using a 
hybrid picture, \ie, superposition of the short-range quark 
exchange interaction and the meson exchange potential at longer 
distances{\cite{BB}}.
Such hybrid models of nuclear force are very successful and furthermore they 
have enough predictive power on the other baryon-baryon interactions.
Recent development of hypernuclear experiments have confirmed some of the 
characteristic features of the QCM based $\Lambda-N$ interactions.

In this short note, I would like to review the status of the QCM and 
describe a new result on the charge symmetry breaking in QCM.

\section{Status of the Quark Cluster Model}

The quark cluster model (QCM) is designed to describe quark exchange 
interactions between baryons.  The basic idea is that the quark 
antisymmetrization  between two three-quark baryons induces symmetry 
interaction, which is either attractive for orbitally symmetric states, 
or repulsive for orbitally antisymmetric states.  Such interaction has 
been well known in the hydrogen molecule as was pointed out by 
Heitler and London in 1927.  It is the symmetry of internal quantum numbers, 
such as spin in the case of molecules, that determines the main features of 
the exchange force.  

In the case of quark exchange interaction, the 
symmetry is controlled by three internal degrees of freedom, color, 
spin and flavor.  Consider the symmetry group for the six quarks, $S_{6}$
in the $BB'$ states, where $B$ and $B'$ are the ground state baryons. 
The color part of the two-baryon state must 
have [222]$_{c}$ symmetry so that the whole system is color \SUc3\ singlet.
As the ground state baryons belong to the 56-dimensional 
or [3] symmetric representation of the spin-flavor \SUsf6, two baryon 
system is classified by 
$[3] \otimes [3] = [6]\oplus [51]\oplus [42] \oplus [33]$
symmetric representations.
Among them, $[6]_{sf}$ and $[42]_{sf}$ are symmetric under 
the exchange of two [3]'s 
and $[51]_{sf}$ and $[33]_{sf}$ are antisymmetric, and therefore
$[51]_{sf}$ and $[33]_{sf}$ are relevant for $L=0$ $BB'$ states.
It is, however, easy to see that the $[51]_{sf}$ symmetric \SUsf6\ state 
cannot have the totally symmetric orbital wave function because
the color $[222]_{c}$ and the spin-flavor $[51]_{sf}$ symmetries 
do not form the totally antisymmetric state:
$ [222]_{c} \times [51]_{sf} \times [6]_{o} \ne [1^6]$
Thus we find that the totally symmetric orbital state such as $(0s)^6$ 
is not allowed for the [51] symmetric \SUsf6\ systems.

Thus the classification of the $BB'$ states in terms of the \SUsf6\ 
representations tells us which $BB'$ acquires strong repulsion at 
short distances.
In the case of $N\Sigma$ interactions, there are four 
possible combinations of spin and isospin.  Among them, 
the ($S=0$, $I=1/2$) and 
($S=1$, $I=3/2$) states should have strong repulsion at short 
distances, because their $S$-wave states are almost forbidden by the Pauli 
principle.  On the contrary, the ``H-dibaryon'' channel, ($S=0$, 
$I=0$) $\Lambda\Lambda-N\Xi-\Sigma\Sigma$, contains only the [33] 
component and therefore ``super-allowed''.  This is one of the reasons 
why ``H-dibaryon'' is expected to have a bound state\cite{H}.
Another reason is that the ``H'' is preferred by the color-magnetic 
interaction.

In the case of $NN$ interactions, the Pauli principle is not enough 
to reproduce the strong short-range repulsion, because the $S$-wave 
$NN$ states consist of only 44\% [51] symmetry forbidden state.
In our original study{\cite{OY}}, we pointed out that the color-magnetic 
part of the gluon exchange interaction is responsible for the repulsion.  
Later it was found that the spin-spin (or hyperfine) interaction is 
the key, whose origin may have several possibilities.
For instance, the instanton induced 
interaction{\cite{T-O}} or the pion exchange interaction{\cite{GR}} may work as well as OgE.
It is also known that the success of the quark model description 
of the meson and baryon spectra owes largely to the same spin-spin 
interaction between quarks. 
For instance, $N-\Delta$ and $\Lambda-\Sigma$ mass 
differences and the negative mean square charge radius of the neutron
are all explained by the spin-spin interaction.
Strange baryons play important roles in confirming the origin of 
short-range interactions.  They include new channels with different 
internal symmetry from $NN$.  Also the $SU(3)$ breaking pattern 
gives further information.

The present status of the QCM can be summarized as follows.
\begin{itemize}
    \item[1.] Nuclear force:  QCM with meson exchange potential 
    supplemented for long range interaction{\cite{BB}} 
    describes $NN$ interaction pretty well.  Main difference among the 
    models appear in the treatment of smooth connection between the 
    short and long range parts.  A natural way is to introduce a form 
    factors that the short-range part of the meson-exchange 
    potential is reduced according to the extension of the baryons, 
    which is about 0.6 fm.  Some disagreement still exists on the 
    origin of the spin-orbit forces among the models.
    \item[2.] $YN$ interaction:  The spin independent part of the 
    central force agrees among the models and with experiment fairly 
    well, while the spin-spin and spin-orbit interactions differ.
    The QCM predicts strong antisymmetric $LS$ force between 
    $\Lambda$ and $N$\cite{ALS}, which is a strong candidate for the small 
    spin-orbit interaction for $\Lambda$ observed in hypernuclei 
    recently.  Strong spin-isospin dependence of the $\Sigma-N$ 
    interaction is yet to be confirmed.
    \item[3.] The QCM predicts a bound state of $\Lambda\Lambda$, or 
    $H$ dibaryon, which was not observed so far.  We found that the 
    instanton induced interaction gives a strong three-quark repulsion 
    to the $H$ state, resulting that the bound state shifts to a 
    resonance above the $\Lambda\Lambda$ threshold\cite{T-O}.  Significant 
    enhancement seen in $\Lambda\Lambda$ final states may be a signal 
    for such a resonance state.
\end{itemize}

\section{Charge Symmetry Breaking (CSB)}

The charge symmetry is a symmetry of QCD provided that 
the $u$ and $d$ quarks have the same mass. 
The symmetry is fairly well satisfied in the hadron spectrum and 
its interactions, although the $u-d$ quark mass (QM) difference  
as well as the electromagnetic (EM) interactions break the symmetry 
by the order of 1\%.
For instance, the mass difference of proton and neutron, 
$$  M_{p}-M_{n} \equiv -\Delta M \simeq -1.3 \hbox{MeV}$$ 
comes from the EM contribution, $\simeq 0.5$ MeV and the 
QM contribution, $\simeq -1.8$ MeV.  Similar mass differences
are observed in the $\Sigma$ and $\Xi$ baryons.

For the two-baryon systems, the charge symmetry predicts the
degeneracy of the $pp$ and $nn$ systems.
After correcting the Coulomb interactions, the scattering lengths of 
$pp$ and $nn$ elastic scatterings show a small difference,
$$ \Delta a \equiv a_{pp} - a_{nn} \simeq 1.5 \hbox{fm}$$
which is attributed to the charge symmetry breaking.
Another interesting observable is the difference of the analyzing 
power in the $p - n$ scattering, $\Delta A \equiv A_{p}-A_{n}$.

For nuclear systems, the charge symmetry relates mirror nuclei, 
a pair of the ($Z$, $N$) and  ($N$, $Z$) nuclei, 
which have similar level structures, while the charge symmetry breaking 
can be observed in the difference in the binding energies, called
the Nolen-Schiffer anomaly.

Various origins of the CSB in nuclear force are discussed in the 
literatures\cite{CSB}.  In the meson exchange picture of the nuclear force, the 
symmetry is broken by the mixing of $\eta$ meson into the pion, or 
the mixing of $\omega$ in $\rho$.  These effects have been studied 
but their effects does not give satisfactory explanation of  
$\Delta a$ and $\Delta A$.  This leads us to consider the CSB in the 
short-range nuclear force.

Thus it is interesting to study how the quark exchange interaction 
contributes to the charge symmetry breaking, which is induced
by the $u$-$d$ quark mass difference, $\Delta m \equiv 
m_{d}-m_{u}$.  
A pioneering work by Chemtob and Yang\cite{CY} showed that the mass 
dependence of the color magnetic gluon exchange gives considerable 
CSB on the $^{1}S_{0}$ $NN$ scattering lengths.
In an extensive analysis of the nuclear binding energies, 
Nakamura \etal{\cite{Nakamura}} pointed out the necessity of 
the short-range CSB, and showed that 
the quark cluster model gives significant contribution which is 
consistent with experimental data.
However, calculation of $\Delta A$ by Br\"auer \etal\cite{Brauer}
concluded that 
the quark exchange contribution is too small for $\Delta A$.

Here I report recent study in collaboration with Nasu and 
Takeuchi\cite{NOT}, in which our aim is to 
describe $\Delta M$, $\Delta a$ and $\Delta A$ simultaneously in the 
quark model picture of the nucleon.
We here assume that the $\Delta m$ for the constituent 
quarks is of the same order as that of the current quark masses,
about 3-5 MeV.  
We consider the potential quark model hamiltonian
\begin{equation}
    H= K + V_{\rm conf} + (1-P_{III})V_{\rm OgE} + P_{III}V_{III}
    + V_{EM}
\end{equation}
Here $V_{\rm conf}$ is the confinement potential, which does not 
distinguish $u$ and $d$ quarks, and therefore charge symmetric,
and $V_{\rm OgE}$ is the one-gluon exchange (OgE) potential, 
whose spin-spin term breaks charge symmetry, \ie,
\begin{equation}
    - \lamlam{\alpha_{s}\over 4}\,\left[
	 {\pi\over m_{i}m_{j}}\, \left(1+{2\over 3} 
	\sigsig\right) \,\delta(\vecr_{ij}) \right]
	\label{eq:Voge}
\end{equation}
$V_{III}$ is the instanton induced interaction, which is effective only for 
the $I=0$ $ud$ quark pair and therefore does not break charge symmetry
nor charge independence.
It is, however, important to consider $V_{III}$ explicitly because
it plays a role to reduce the effect of OgE.
$V_{EM}$ is the Coulomb interaction.  The kinetic energy term of 
quarks depends on the quark masses and therefore contains CSB.

Our aim is to confirm that the quark mass difference in the 
constituent quark picture gives consistent magnitudes of all three 
experimental data, $\Delta M$, $\Delta a$ and $\Delta A$.
To keep the consistency, we keep our evaluation of the CSB effect up to 
the lowest order in the perturbation theory throughout our work.  In 
the scattering processes, we evaluate the distorted-wave Born term 
only and the difference of the $T$-matrix is used to evaluate 
$\Delta a$ and $\Delta A$.

I cannot go into the detail in this short report, but the summary of 
our results is given here.  Please refer to {\cite{NOT}} for details.
\begin{enumerate}
    \item[1.] The mass difference of proton and neutron requires the 
    introduction of $III$, which gives hyperfine splitting of the baryon 
    spectrum but does not induce CSB.  The reason for the importance 
    of $III$ is that the color magnetic interaction of the gluon 
    exchange gives  too strong CSB in the wrong direction so that the 
    quark mass difference is essentially cancelled.  After 
    introducing $III$, which shares more than 40\% of the hyperfine 
    splitting by one gluon exchange interaction, we can explain 
    $\Delta M$. We have determined $\Delta m$ and $P_{III}$ 
    so that the neutron-proton mass difference is reproduced (Table 1).
    \item[2.] The $^{1}{\rm S}_{0}$ scattering length has been 
    evaluated in the quark cluster model and the results are 
    summarized in Table 1.  We find that a large contribution comes from 
    the kinematical effect that is induced by the mass difference of 
    the proton and neutron.  Note that the scattering length is inversely 
    proportional to the derivative of the T matrix in terms of the 
    wave number $k$, the baryon mass difference, which changes the 
    dispersion relation should contribute.
    The results show that the reduced OgE can also explain $\Delta a$.
    \item[3.]  The analyzing powers are calculated by summing the 
    $L=0, 1, 2$ waves.  The antisymmetric $LS$ force is the source of 
    CSB,
    $$   V_{\rm eff CSB} \propto (\sigv_{A}-\sigv_{B})\cdot\vecL_{AB} 
    \times (\tau_{3}^{A}-\tau_{3}^{B}) $$
    This interaction induces couplings of $^{3}{\rm P}_{1}$ and
    $^{1}{\rm P}_{1}$, and $^{3}{\rm D}_{2}$ and $^{1}{\rm D}_{2}$,
    or in general $^{3}{\rm L}_{L}$ and $^{1}{\rm L}_{L}$.  In the 
    case of $NN$ scattering these mixings break isospin invariance 
    and charge symmetry.  Thus the differences of the analyzing power 
    of the proton and that of the neutron in the $pn$ scattering are
    direct CSB contribution.  Fig.~1 shows the prediction of $\Delta 
    A$ in our QCM calculation, compared with the data from Indiana.
    The results seem consistent with the data.
\end{enumerate}

\begin{table}[htb]
    \caption{The $^{1}{\rm S}_{0}$ $NN$ scattering length, effective 
    range parameters and their CSB effects. $\Delta m$ and $P_{III}$ 
    are determined so that $\Delta M \simeq 1.3$ MeV. 
    The other quark model parameters are fixed by the baryon spectrum:
    $\bar m_{q}= 313$ MeV, $b=0.6$ fm, $\alpha_{s}=1.52$, and 
    $a_{\rm conf} =60.1$ MeV/fm.}
    \begin{center}
	\begin{tabular}{c|cccccc} \hline 
	    & $P_{III}$ & $\Delta m$ [MeV] & $\bar{a}$ [fm] & $\bar{r}$ [fm] 
	    & $\Delta a$[fm] & $\Delta r$ [fm] 
	    \\ \hline 
	    A & 0.4 & 7.625 & -17.9 & 2.43 & 1.4 & -1.0  \\ 
	    B & 0.5 & 5.375 & -17.9 & 2.46 & 1.1 & -0.8  \\ 
	    Observed$^{a} $ &  &  & -18.1$\pm$0.4 & 2.80$\pm$0.11 
	    & 1.5$\pm$0.5 & 0.1$\pm$0.12 \\ 
	    \hline
	\end{tabular}
    \end{center}
\end{table}

\begin{figure}
       \centerline{\includegraphics[width=11 cm]
                                   {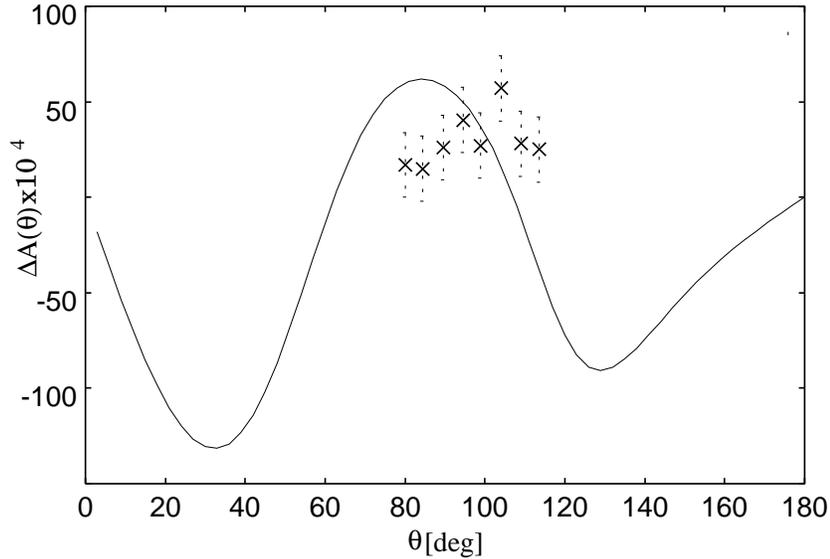}}
   \caption{$\Delta A$ at 183 MeV $pn$ scattering.}
   \label{Delta A}
\end{figure}

It should be noted that all the quark model parameters including the 
CSB interactions are determined in the baryon spectrum in our study 
and no CSB is induced from the other elements.  This is in a sense an 
extreme because the meson exchange interaction also causes certain 
amount of the CSB.  It is, however, confirmed that  the quark mass 
difference the quark cluster model approach gives significant and 
consistent amount of the CSB.

\section{Conclusion}

The quark cluster model is classic but it still has practical power 
in describing baryon-baryon interaction.  The success of QCM largely 
owes to the special role of the strangeness.  
Recent development in hypernuclear physics, in which 
interactions of strange baryons are studied extensively, provided us 
with qualitatively new information on the baryon-baryon interactions.
Further development is expected to make the mechanism of the 
baryon-baryon interactions much clearer than our present knowledge.

In this talk, I have discussed the charge symmetry breaking in the 
$NN$ interaction from the QCM viewpoint and have shown that a 
consistent picture can be drawn by assuming the CSB comes from the 
short range part of the interactions.  It is yet to be concluded 
how the long-range and short-range parts share the roles in CSB in 
nuclear force, but it is significant that the contribution of the 
short-range part is important.

\bigskip

The main part of this work has been done in collaboration with
Takashi Nasu and Sachiko Takeuchi.

\end{document}

%% file: defs.tex
\def \bracket<#1>{\mbox{$\langle {#1}\rangle$}}
\def \brav #1|{\mbox{$\langle {#1}|$}}
\def \cg(#1,#2,#3,#4,#5,#6){\mbox{$(#1,#2,#3,#4|#5,#6)$}}
\def \comm[#1,#2]{\mbox{$\left[{#1},{#2}\right]$}}
\def \ddt #1{\ifmmode{{\partial #1\over\partial t}} 
\else{${\partial #1\over\partial t}$}\fi}

\def \dotp(#1.#2){\mbox{$(#1\cdot #2)$}}

\def \ketv #1>{\mbox{$|{#1}\rangle$}} 
\def \mate<#1|#2|#3>{\mbox{$\langle {#1}|\,{#2}\,|{#3}\rangle$}}
\def \mated<#1||#2||#3>{\mbox{$\langle {#1}||\,{#2}\,||{#3}\rangle$}}
\def \mvec #1{\mbox{\boldmath{${#1}$}}}
\def \rtov(#1/#2){{\ifmmode{{\sqrt{\frac{#1}{#2}}}} 
\else{{$\sqrt{\frac{#1}{#2}}$}}\fi}}
\def \sixj(#1,#2,#3,#4,#5,#6)
{\mbox{$\left\{\matrix {#1&#2&#3\cr#4&#5&#6\cr}\right\}$}}

\def\SUsf#1{\mbox{$SU(#1)_{sf}$}}
\def\SUc#1{\mbox{$SU(#1)_{c}$}}
\def \Trace[#1]{\mbox{${\hbox{Tr} \left\{#1\right\}}$}}
\def \xp(#1.#2){\mbox{${(#1\times #2)}$}}

\def \etal{{\it et al.}}

\def \ie{{\it i.e.}}

\def \lam{\lambda}
 
\def \lamlam{\mbox{$(\lam_i\cdot\lam_j)$}}

\def \lam{\mbox{$\lambda$}}

\def \rhat{\ifmmode{\hat{\mvec r}}\else{$\hat{\mvec r}$}\fi}

\def \sigsig{\mbox{$(\sigv_i\cdot\sigv_j)$}}
\def \sigv{\mbox{$\vecsigma$}}

\def \vecL{\mvec L}

\def \vecr{\mvec r}
\def \vecsigma{\mvec \sigma}

\def \ddt #1{\ifmmode{{\partial #1\over\partial t}} 
\else{${\partial #1\over\partial t}$}\fi}

\newif\ifrefphysrev \newif\ifrefPTP

\def\refPTP{\refphysrevfalse\refPTPtrue
           \typeout{** Reference: Progress format}}
\refPTP

\def \vol(#1,#2,#3){\ifrefphysrev{{\bf {#1}}, {#3} (19{#2})}%
\else{\ifrefPTP{{\bf {#1}} (19{#2}), {#3}}%
\else{{\bf {#1}}(19{#2}){#3}}\fi}\fi}
\def \NP(#1,#2,#3){Nucl.\ Phys.\          \vol(#1,#2,#3)}
\def \PL(#1,#2,#3){Phys.\ Lett.\          \vol(#1,#2,#3)}
\def \PRL(#1,#2,#3){Phys.\ Rev.\ Lett.\   \vol(#1,#2,#3)}
\def \PRp(#1,#2,#3){Phys.\ Rep.\          \vol(#1,#2,#3)}
\def \PR(#1,#2,#3){Phys.\ Rev.\           \vol(#1,#2,#3)}
\def\PRD(#1,#2,#3){\PR(D#1,#2,#3)}
\def\PRC(#1,#2,#3){\PR(C#1,#2,#3)}
\def \PTP(#1,#2,#3){Prog.\ Theor.\ Phys.\ \vol(#1,#2,#3)}
\def \PTPS(#1,#2,#3){Prog.\ Theor.\ Phys.\ Suppl.\ \vol(#1,#2,#3)}
\def \ibid(#1,#2,#3){{\it ibid.}\         \vol(#1,#2,#3)}

\def \ST{S.\ Takeuchi}
\def \KY{K.\ Yazaki}
\def \MO{M.\ Oka}
\def\KS{K.\ Shimizu}\def\KShimizu{\KS}